\documentstyle[prl,aps]{revtex}
\begin{document}
% ------------------------------------------------------------------------------
\title{Tunneling and Universality in the Integer Quantum Hall Effect} 
\author{Alex Hansen\footnote{Permanent address: Department of Physics,
Norwegian University of Science and Technology, NTNU, N--7034 Trondheim, 
Norway. E-mail: alex.hansen@phys.ntnu.no}}
\address{Department of Theoretical Physics, University of Cologne,
Z{\"u}lpicherstr.\ 77, D--50937 Cologne, Germany}
\author{J\'anos Kert{\'e}sz\footnote{E-mail: kertesz@planck.phy.bme.hu}}
\address{Department of Theoretical Physics, Institute of Physics, Technical 
University of Budapest, H--1111 Budapest, Hungary}
\date{June 5, 1997}
\maketitle
% --------------------------------------------------------------------
%  ABSTRACT
%---------------------------------------------------------------------
\begin{abstract}
We show analytically and numerically that omission of quantum
interference from the Chalker--Coddington model of the integer
quantum Hall effect gives a localization length exponent $\nu=4/3$ as in
ordinary two-dimensional percolation.  Thus, contrary to semi-classical
scaling arguments, tunneling alone does not lead to $\nu=7/3$. 
\end{abstract}
% --------------------------------------------------------------------
% --------------------------------------------------------------------
%   MAIN TEXT
% --------------------------------------------------------------------
Scaling and universality in the integer quantum Hall effect (IQHE)
related to the localization--delocalization
transition of charge carriers when going from one plateau to the next
is of considerable interest \cite{pg87,jvfh94}.
By measuring how the width of the peaks of the longitudinal conductance scales
with magnetic field $B$ and linear size $W$ of the system, Koch 
et al.\ \cite{khkp91} found the scaling law $W\sim |B-B_c|^{-\nu}$ where 
$\nu\approx 2.3$ and $B_c$ is the value of the magnetic field needed to
have half filling.

This value excludes early theories\cite{kl82,t83} that predicted
universal two-dimensional percolation exponent ($\nu = \nu_p = 4/3$).
However, much before the measurements, Mil'nikov and Sokolov
\cite{ms88} argued on the basis of a semi-classical picture and
scaling that the classical value should be shifted by unity, i.e.\
$\nu=7/3$.  Numerically 2.35 was found \cite{cc88,h95}
in the Chalker--Coddington (CC) model \cite{cc88}, which includes in
addition to the semi-classical tunneling effect also interference, and
seems to capture the critical behavior in IQHE. It has been unclear
whether tunneling is sufficient to explain, at least in a range of
parameters, the change of the localization length exponent from the
classical percolation value, 4/3, to the observed one, ($\sim 7/3$)
\cite{hk90,feng93,hl95}.

We present here a study of the effect of tunneling without interference in
a CC type model.  Our conclusions, based on analytic arguments and 
tested numerically, are as follows: 1) Tunneling alone does {\it not\/} 
change the localization length exponent from its classical percolation value,
4/3.  2) The value actually 
observed is caused by the mechanism we left out (but emphasized by CC):
Quantum interference.

In the classical picture the delocalization of the
charge carriers is considered as a percolation transition \cite{kl82,t83}.  
Noninteracting charges are assumed to move in a disordered potential 
slowly varying on the scale of the magnetic length 
$l=\sqrt{\hbar /eB}$ (high field model).  In this
approximation, the electron wave function $\psi$ is localized to 
equipotential curves, forming ``ribbons" of width $l$.  Changing $B$ causes 
the localized wave function to move up or down in this landscape.  
Typically, the ribbons constituting the wave function are restricted to 
encircling mountains or valleys in the energy landscape. However, scattering 
between different ribbons is possible when they overlap at saddle points.  
There is a typical distance $a$ between saddle points.  In a window 
$\Delta B$ centered at $B_c$, a sample-spanning cluster of ribbons 
overlapping at saddle points exists, resulting in delocalization.  The width 
of the window $\Delta B$ is determined by the magnetic length $l$ and
decreases with increasing sample size $W$.
This classical description becomes semi-classical once the
question of overlap of wave functions across saddle points takes
tunneling into account.  Overlap is then characterized by a tunneling
probability \cite{f88}
\begin{equation}
P={1\over{1+e^{1/\chi}}}\  \ \ ;
\ \ \ \chi={{ml^2|V_{xx}V_{yy}|^{1/2}}\over{h e|B-B_c|}}\;.
\label{eq1}
\end{equation}
Here $V_{xx}$ and $V_{yy}$ are the principal double spatial derivatives of
the disordered potential at the saddle point, and $m$ is the effective 
electron mass.  The semi-classical argument leading to $\nu=\nu_p+1=7/3$ 
is essentially based on a change of effective basic length scale of the 
percolation problem from $a$ to $\chi a$.  However, there is an underlying
assumption that the tunneling does not change in any essential way the
structure of the infinite cluster defined by the topology of the paths
in the random energy landscape.  As we will show later, this is the
weak part of the argument, and the cause of its breakdown.

In order to proceed we study the semi-classical argument using the
extended CC model \cite{lwk93}. We define first a fully
classical (percolation) version \cite{rsg88}. Imagine a square lattice
of linear size $L$ where every plaquette is occupied by one of the
tiles of Fig.\ \ref{fig1}.  The result is a set of lines which are either
closed or end at the boundary of the sample,
see Fig.\ \ref{fig2}a. A control parameter $p$ tells the 
probability of the tile 1a. This problem can be mapped exactly to bond
percolation on a square lattice \cite{rsg88}.  The mapping is accomplished 
by drawing a diagonal on each tile oriented in such a way that it does not
intersect the two quarter circles already on the tile. The resulting pattern 
of diagonals is shown in Fig.\ \ref{fig2}a.  This
pattern consists of two  lattices oriented $45^\circ$ with respect to the 
original one defined by the tiles, and with lattice constant equal to the 
length of the diagonals of the tiles.  The two lattices are displaced by a
distance equal to the linear size of the tiles in the two principal orthogonal
directions.  Furthermore, whenever there is a bond present in one of the  
two lattices, the corresponding one is missing in the other lattice, and 
{\it vice versa,\/} making the two lattices duals of each other. This is 
shown in Figs.\ \ref{fig2}a, b and c. This makes the problem of random 
orienting the two tiles of Fig.\ \ref{fig1} equivalent to placing a bond
or not in one of the two intertwined lattices, say that of
fig. \ref{fig2}b.
It is also clear that, for $p=1/2$  we deal with a percolation problem
at the critical point, as with equal probability in choosing either tile
corresponds to placing a bond with probability $1/2$, which is the 
percolation threshold on the square lattice. 

The bonds of the tiling represent the classical paths the electrons
may take in the disordered energy landscape. Each bond can only be
traversed in one direction.  The possible directions are chosen so
that the network forms a grid of loops of alternating handedness
according to the magnetic field and the surface topology.  The centers
of the the tiles 
represent saddle points, and the lattice constant corresponds to $a$.
The question of delocalization is equivalent to whether there is a
path spanning the system.

We now introduce tunneling within the framework of a semi-classical 
approximation. The transmission probabilities at a given saddle point are 
controlled by a parameter $\gamma $ characterizing the amount of
tunneling \cite{lwk93}.  By adjusting it to the usual control 
parameter of the CC model determining the scattering properties at the 
saddle point we have
\begin{equation}
\chi={1\over{2|\ln\sinh \gamma |}}\;.
\label{eq2}
\end{equation}
When $\gamma\to 0$ or $\gamma\to\infty$, tunneling is negligible, while when 
$\gamma=\gamma_c=\ln(1+\sqrt{2})$, the transmission probability is 1/2.

Following Lee et al.\ \cite{lwk93}, we assume $\gamma=\gamma_c e^{\mu-v}$. 
Here $\mu$ is the dimensionless chemical potential equal for all saddle
points in the network, while 
$v$ is chosen randomly on the interval $[-w/2,+w/2]$ representing the
random potential.  In the case when 
$|\mu-v|\ll 1$, we may relate $|\mu-v|\propto |B-B_c|$ of the 
high-field model. In extreme case when $w\to\infty$, there is no 
contact across the saddle points and the model is just the classical version 
discussed above with 1/2 occupation probability for both tiles.  
Delocalization happens then as a classical percolation phenomenon and, 
accordingly, the correlation length exponent has to be $4/3$.

We now turn to the case of finite $w$ when tunneling becomes
important. In the CC model one should assign phases to the electrons
moving in the channels and the change of the phases during tunneling
should be taken into account. We neglect the phases here since the aim
of this work is to test the effect of tunneling {\it alone\/}. What is the action of
tunneling near a given saddle point, i.e.\ on a given tile in this
Fig.\ \ref{fig2}a? It simply means that, with some probability, a tiling in
the classical picture has 
to be changed to the other one for this particular electronic process.
This corresponds in the bond percolation picture either to remove or
to place a bond with some well defined probability that mimics
Eqs.\ (\ref{eq1}). Thus, instead of introducing a $\gamma=\gamma_c
\exp(\mu-v)$ and relating it to the tunnel probability $P$ by Eq.\
(\ref{eq2}), we can divide the assigning construction of the
conducting paths into two parts: one which places the ``classical"
paths, and then a second probability that models the tunneling.  Given
a control parameter $p\in [0,1]$, we draw two random numbers $r$ and
$\rho$ from a uniform distribution between zero and one for each bond.
The bond is present if the following function is equal to one,
\begin{equation}
\label{eq3}
\pi(p, \Lambda ;r,\rho)=\theta(p-r)
+\left(\theta(r-p)-\theta(p-r)\right)\theta(P-\rho)\;,
\end{equation}
where, $\theta$ is the step function and $P$ is defined in Eq.\
(\ref{eq1}), 
with $1/\chi=\Lambda |p-r|$ and 
$\Lambda$ being a scale factor. In words: Tunneling introduces a bond if
there was no one classically and 
cancels the bond if it was present.  This means that actual
realizations of paths are followed and an average over such paths
should finally carried out.

The smaller $\Lambda$, the more important is tunneling. The
idea behind Eq.\ (\ref{eq3}) is simple. If $\Lambda\to\infty$,
$P\to 0$, and Eq.\ (\ref{eq3}) reduces to
$\pi(p;r,\rho)=\theta(p-r)$, which is describing a simple bond
percolation problem with $p$ as control parameter.  We may therefore
interpret $r$ as characterizing the level of the corresponding saddle point, 
and $p$ the magnetic field.  The second term in Eq.\ (\ref{eq3}) will
switch the value of $\pi$ given by the first ``classical" term to
the opposite value with a probability which falls off exponentially
the further $p$ is away from the value $r$.  In particular, when $p=r$
for bond $i$, there is a 50\% chance that the value of $\pi$ will be
reversed compared to its classical value.  Furthermore, the further
$r$ and $p$ are apart, the more difficult to switch the value of
$\pi$.  This completes the definition of our tunnel-only version of
the Chalker-Coddington model.

The probability that a bond is present is then 
\begin{equation}
\label{eq5}
\overline{\pi}(p)=
\langle \pi(p;r,\rho)\rangle=\int_0^1 dr \int_0^1 d\rho\ \pi(p;r,\rho)
=p+\frac{1}{\Lambda}
\log\left(\frac{1+e^{-\Lambda p}}{1+e^{-\Lambda (1-p)}}\right)\;,
\end{equation}

Since assigning a bond or not is a purely local decision, which is so far
based on two random numbers $r$ and $\rho$, we may simplify and draw a
single random number $q$ between zero and one and comparing this to 
$\overline{\pi}$: 
\begin{equation}
\label{eq6}
\pi(p;q)=\theta(\overline{\pi}-q)\;.
\end{equation}
This formulation is equivalent to that of Eq.\ (\ref{eq3}).  It is 
simpler to analyze, but harder to interpret.  The effect of tunneling 
has been to produce a non-linear relationship between the control parameter
$p$ and the probability for a bond to be present.  However, we have
that $\overline{\pi}=1/2$ when $p=1/2$, as expected: The system
has the same critical point with or without tunneling: $p_c=1/2$.

We are now ready to analyze what happens close to the critical value of
$p$, and in particular determining the localization length exponent 
$\nu$ in this model.  In order to do this, we first note that using
$\overline{\pi}$ as control parameter and not $p$, makes the problem a
standard percolation problem.  Thus, the correlation length $\xi$ diverges
as
\begin{equation}
\label{eq7}
\xi\sim |\overline{\pi}_c-\overline{\pi}|^{-\nu_p}\;,
\end{equation}
where $\overline{\pi}_c=1/2$, and $\nu_p=4/3$.  We now expand 
$\overline{\pi}$ in powers of $\delta p=p_c-p$ to get
\begin{equation}
\label{eq8}
\overline{\pi}(1/2+\delta p)=1/2 +\tanh\left(\frac{\Lambda}{4}\right)
\delta p + {\cal O}(\delta p^3)\;.
\end{equation}
Plugging this expansion into Eq.\ (\ref{eq8}) gives
\begin{equation}
\label{eq9}
\xi\sim |p_c-p|^{-\nu_p}\;,
\end{equation}
but with a different prefactor.  The exponent in this equation is by
definition $\nu$ and we have shown that 
\begin{equation}
\label{eq10}
\nu=\nu_p=\frac{4}{3}
\end{equation}
in the CC model when only tunneling is taken into account.
Based on this result, we conclude that the semi-classical argument giving
$\nu=\nu_p+1$ fails.

It is interesting to note that in a one-dimensional version of this model,
first studied in \cite{hl95}, there {\it is\/} a shift of $\nu$ in
comparison to $\nu_p=1$ by precisely one \cite{hk97}.  
Thus, in this case, the
semiclassical argument {\it does\/} work. The difference between the one and two-dimensional cases is that in the 
latter the number of bonds almost belonging to the infinite cluster is so
large --- their density is finite --- that the system simply is driven
away from criticality.  

We have carried out numerical simulations of the semi-classical 
CC model as defined above in order to see the way 
asymptotics sets in. We generated different systems sizes up to 
$1100\times 1100$ with at least 20 000 samples using Monte Carlo 
renormalization group \cite{rsk80} to determine the exponents. We have 
based our simulations on Eq.\ (\ref{eq6}) directly, and on Eq.\ 
(\ref{eq3}) directly.  In the latter case, we used {\it geometrical\/} sample
averaging in order to test whether the {\it typical\/} percolation 
probability behaves differently from the average one.  We found no 
difference in the critical behavior.  The measured effective exponent was
$\nu=1.53$ for $W=100$ and $\nu=1.45$ for $W=1100$. Thus there is a
slow convergence towards the predicted universal exponent $4/3$ but
even small size samples have an effective exponent much smaller than
$7/3$, confirming our analytical conclusion.

The localization length exponent  observed in the 
Chalker-Coddington model and in other models of the integer quantum Hall
effect in addition to experiments, is larger than $\nu_p$ by about unity.
What we have shown here is that the tunneling mechanism alone is not
able to account for the change of the exponent in any range of
the parameters -- therefore interference must be
responsible for it.

As the control parameter $\Delta B$ increases the interference becomes
less and less pronounced and the entangled paths due to tunneling
percolation dominate. This should be reflected in a crossover from the
quantum exponent $\nu \approx 2.3$ to the classical value $4/3$. We
suggest the experimental check of this predition.

\vskip 0.5cm
We thank J.\ Hajdu, E.H. Hauge, J.\ Hove, B.\ Huckestein, R.\ Klesse,
M. Metzler and D. Polyakov for 
valuable discussions.  A.H.\ furthermore thanks J.\ Hajdu for 
the invitation to Cologne where part of this work was done. Partial
support by OTKA T016568 (JK) and SFB341 (AH) is acknowledged.
% --------------------------------------------------------------------
% BIBLIOGRAPHY
% --------------------------------------------------------------------

% --------------------------------------------------------------------
% FIGURE CAPTIONS
% --------------------------------------------------------------------
\begin{figure}
\caption{\label{fig1} The two fundamental tiles constituting the
building blocks of the CC model in the classical limit.}
\end{figure}
\begin{figure}
\caption{\label{fig2} (a) A set of dragon curves with corresponding diagonals
inscribed for $p=1/2$.  In (b) and (c) we have decomposed the set of 
diagonals into a sublattice (b) and the corresponding dual lattice (c).}
\end{figure}
% --------------------------------------------------------------------
\end{document}